\documentstyle[sprocl,epsf]{article}

\def\Journal#1#2#3#4{{#1} {\bf #2}, #3 (#4)}

\def\NPB{{\em Nucl.\ Phys.} B}
\def\PLB{{\em Phys.\ Lett.}  B}

\def\PRD{{\em Phys.\ Rev.} D}
\def\AP{{\em Ann.\ Phys.}}

\def\JPA{{\em J.\ Phys.} A}

\def\be{\begin{equation}}
\def\ee{\end{equation}}
\def\bea{\begin{eqnarray}}
\def\eea{\end{eqnarray}}

\def\beeq{\begin{equation}}
\def\eneq{\end{equation}}
\def\beqn{\begin{eqnarray}}
\def\eeqn{\end{eqnarray}}

\def\mybig{\displaystyle \strut }

\def\dd{\partial}
\def\la{\raise.16ex\hbox{$\langle$}\lower.16ex\hbox{}  }
\def\ra{\, \raise.16ex\hbox{$\rangle$}\lower.16ex\hbox{} }
\def\go{\rightarrow}

\def\onehalf{ \hbox{${1\over 2}$} }

\def\psibar{ \psi \kern-.65em\raise.6em\hbox{$-$} }
\def\psibaralpha{ \psi^{(\alpha)} \kern-1.9em\raise.6em\hbox{$-$}
\kern+1.2em\hbox{}}
\def\chibar{ \chi \kern-.65em\raise.5em\hbox{$-$} }
\def\mbar{ m \kern-.75em\raise.4em\hbox{$-$}\hbox{} }
\def\Bbar{ B \kern-.73em\raise.6em\hbox{$-$}\hbox{} }

\def\L{ {\cal L} }
\def\ep{\epsilon}

\def\wil{ \Theta_{\rm W} }

\def\vphi{ {\varphi} }

\def\eff{{\rm eff}}

\def\myfrac#1#2{{\mybig #1 \over \mybig #2}}
\def\myitem#1{{\noindent {\hglue 5pt}#1{\hglue 5pt}}}

\def\boxit#1{$\vcenter{\hrule\hbox{\vrule\kern3pt
     \vbox{\kern3pt\hbox{#1}\kern3pt}\kern3pt\vrule}\hrule}$}
\def\bigbox#1{$\vcenter{\hrule\hbox{\vrule\kern5pt
     \vbox{\kern5pt\hbox{#1}\kern5pt}\kern5pt\vrule}\hrule}$}
\def\hugebox#1{$\vcenter{\hrule\hbox{\vrule\kern8pt
     \vbox{\kern8pt\hbox{#1}\kern8pt}\kern8pt\vrule}\hrule}$}

\begin{document}

\rightline{\small UMN-TH-1431/96}
\vglue .7cm


\centerline{\bf GAUGE~ THEORY~ MODEL \footnote{~To appear 
in the Proceedings of {\it the Second International Sakharov
Conference on Physics}, Lebedev Physical Institute, Moscow, May 20-24,
1996.}}

\title{-- Quark Dynamics and Anti-ferromagnets --}



\author{YUTAKA HOSOTANI}

\address{School of Physics and Astronomy, University of Minnesota\\
Minneapolis, MN 55455, USA}


\maketitle\abstracts{
Two-dimensional QED with N-flavor fermions serves as a model of 
quark dynamics in  QCD as well as an effective theory of an
anti-ferromagnetic spin chain.
It is reduced to N-degree quantum mechanics in which a potential
is self-consistently determined by the Schr\"odinger equation
itself.}

\section*{QCD and antiferromagnets}

There are three reasons to investigate two-dimensional QED with
massive fermions: \cite{HHI1}
\beeq
{\cal L} = - \hbox{$1\over 4$} \, F_{\mu\nu} F^{\mu\nu} + 
\sum_{a=1}^N \psibar_a \Big\{ \gamma^\mu (i \dd_\mu - e A_\mu) -
  m_a  \Big\} \psi_a ~~~.
\label{qed2}
\eneq
First of all, it carries many features
essential in QCD;  confinement, chiral condensates, and $\theta$
vacuum.  One can evaluate various physical quantities such as 
chiral condensates, Polyakov loop, and string tension at zero and finite 
temperature with arbitrary fermion masses to explore QCD physics.

Secondly QED is an effective theory of spin systems.\cite{Itoi}
A $s=\onehalf$ anti-ferromagnetic spin chain 
\beeq
H = J \sum \vec S_n \cdot \vec S_{n+1}  \hskip 1cm
(S=\onehalf, J>0)
\label{spinchain}
\eneq
is equivalent to two flavor massless QED$_2$ in a uniform charge
background in the strong coupling limit.  Similarly  a  spin ladder
system is equivalent to  coupled QED$_2$.   Their correlation functions can
be evaluated systematically.  Physics of two-dimensional
anti-ferrromagnets is the core to understand  the  high
$T_c$ superconductivity, for which QED description provides an 
indispensable tool.

Thirdly there is significant development in technology.
We show that a field theory problem is reduced 
to a quantum mechanics problem of finite degrees of freedom, which
can be solved numerically on work stations.  Technically this method
is much simpler and easier to handle than the lattice gauge theory 
or light front method.

\section*{Reduction to quantum mechanics}

Consider the model (\ref{qed2}) defined on a circle $S^1$ with a
circumference
$L$. With periodic and anti-periodic boundary conditions imposed
on bosonic and fermionic fields,   the model is mathematically 
equivalent to a theory defined on a line ($R^1$) at finite temperature. 
Hence various physical quantities at $T\not=0$ on $R^1$ are obtained from
the corresponding ones at $T=0$ on $S^1$ by substituting  $L$ by
$T^{-1}$.

$\psi_a$ is bosonized on a circle.  It is expressed in terms of
zero modes $q_a$ and oscillatory mode $\phi_a(x)$.
The only physical degree of freedom associated with gauge fields is the  
Wilson line phase $\wil$ along the circle.

When fermions are massless, zero modes ($q_a, \wil$) decouple from
oscilatory modes $\phi_a$.  The latter consists of 1 massive boson
and $N-1$ massless bosons.  The model is exactly solvable.

Fermion masses provide nontrivial interactions among zero and oscillatory
modes. All boson fields become massive.
When  fermion masses
are degenerate, $m_a=m$, we have 1 heavy
boson with mass $\mu_1$ and $N-1$ lighter bosons with mass $\mu_2$.
The vacuum wave
function is  written as $f(p_W,\vphi_1, \cdots, \vphi_{N-1};\theta)$.
$\theta$ is the  vacuum angle parameter of the theory.
$f(p_W,\vphi)$ must satisfy
\beqn
&&\big\{ K + V \big\} ~ f = \ep ~ f  \cr
\noalign{\kern 7pt}
&&K =  -  {N^2\over 4\pi^2}  {\dd^2\over \dd p_W^2}  
- (N-1) \bigg\{ 
 \sum_{a=1}^{N-1} {\dd^2\over \dd\vphi_a^2} 
-{2\over N-1} \sum_{a<b}^{N-1} {\dd^2\over \dd\vphi_a\dd\vphi_b}
\bigg\} \cr
\noalign{\kern 4pt}
&&V(p_W,\vphi)={(\mu Lp_W)^2\over 4} 
 - {NmL \Bbar\over \pi}
\sum_{a=1}^N    \cos \Big( \vphi_a - {2\pi p_W\over N}\Big) 
\label{Schro1}
\eeqn
where  $\sum_{a=1}^N \vphi_a= \theta$, $\mu^2=Ne^2/\pi$  and
$\Bbar= B(\mu_1 L)^{1/N} B(\mu_2 L)^{(N-1)/N}$.  $B(z)$ is given by
$B(z) = (z/4\pi)e^{ \gamma + (\pi/ z)} \exp \big\{
- 2 \int_0^\infty dx / (e^{z\cosh x} -1)  \big\}$.

The boson masses are determined by 
\beeq
\mu_1^2 = \mu^2 + \mu_2^2  ~~~,~~~
\mu_2^2 = {8\pi  m\Bbar \over L}
\la \cos \Big(\vphi_a - {2\pi p_W\over N} \Big) \ra_f
\label{bosonmass}
\eneq
where $\la ~ \ra_f$ denotes the $f$ average.  (\ref{Schro1}) and
(\ref{bosonmass}) are solved simultaneously.   Schematically
$V(p_W,\vphi) \go f(p_W,\vphi) \go \mu_\alpha \go V(p_W,\vphi)$.
This is a
Schr\"odinger problem in which the potential has to be reproduced by the
equation itself.

\section*{Quark dynamics}

Chiral condensates are given by
\beeq
\la \psibar_a\psi_a \ra_\theta = - {2\Bbar\over L}
\la \cos \Big(\vphi_a - {2\pi p_W\over N} \Big) \ra_f~~~.
\label{condensate}
\eneq
The string tension $\sigma$ between two external sources, one with
charge $+q$ and the other with $-q$, is
\beeq
\sigma = Nm \Big\{ \la \psibar\psi\ra_{\theta_\eff}
 - \la\psibar\psi\ra_\theta \Big\} ~~~,~~~
\theta_\eff = \theta-\myfrac{2\pi q}{e} ~. 
\label{string1}
\eneq
These quantities are evaluated numerically with arbitrary
values for $L=T^{-1}$, $m$, and $\theta$.  At $T=0$
and for $m\ll \mu$, 
\beeq
{\sigma \over \mu^2}  = -  {N\over 2\pi} \, \Big( 2e^\gamma
\, {m\over \mu} \Big)^{{2N\over N+1}} 
\bigg\{ \Big( \cos {\bar\theta_\eff\over N}  \Big)^{{2N\over N+1}}
-\Big( \cos {\bar\theta\over N}  \Big)^{{2N\over N+1}} \bigg\}
\label{string2}
\eneq
where 
$\bar\theta=\theta-2\pi[(\theta+\pi)/2\pi]$.  The $q$ dependence of
the string tension at various temperature is displayed in fig.\ 1.

\begin{figure}[t,b]
\epsfxsize= 9.cm    
\epsffile[100 240 450 500]{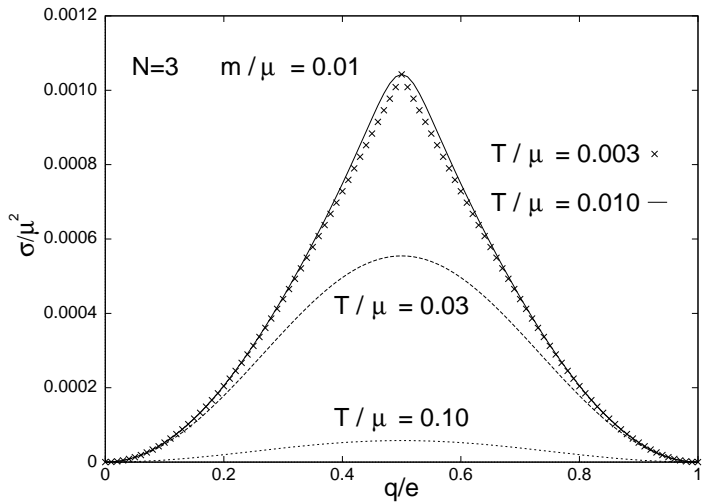}
\caption{The charge, $q$, dependence of the string tension in
the $N=3$ model with $m/\mu=.01$ at $\theta=0$ at various
temperature $T$.  At $T=0$ a cusp singularity develops at
$q=\onehalf e$.
\label{fig:s-tension}}
\end{figure}

The following conclusions are obtained.

\vskip 6pt

\myitem{1.}At  $T=0$ the chiral condensate is not analytic in
  $m$. \cite{Coleman}

\myitem{2.}At $T=0$, there appears a cusp singularity at
$\theta=\pi$.

\myitem{3.}Sufficiently large asymmetry in fermion masses removes
the cusp singularity at $\theta=\pi$.

\myitem{4.}Witten's picture \cite{Witten} of chiral dynamics in QCD$_4$
is reproduced in QED$_2$.

\myitem{5.}The string tension vanishes for an integer $q/e$. \cite{CJS}

\myitem{6.}The string tension  is non-vanishing  only when $m\la
\psibar\psi\ra$ has non-trivial $\theta$ dependence.

\myitem{7.}The chiral condensate increases as a fermion mass becomes
very large: $\la \psibar\psi\ra_{T=0} \sim - (e^{2\gamma}/\pi)\, m$ for $m \gg
\mu$.

\myitem{8.}However, a contribution of a heavy fermion to the string tension
becomes negligible as its mass becomes large.

\myitem{9.}At $\theta=\pi$ a discontinuity in chiral condensates develops
at a critical fermion mass $m_c$.  

\section*{Gauge theory of anti-ferromagnetic spin chains}

A $s=\onehalf$ spin chain, (\ref{spinchain}), is equivalent to QED$_2$.
The derivation goes as follows.

Write $\vec S_n = c_n^\dagger \onehalf \vec \sigma c_n^{}$ where
$c_{n\alpha}$ is an annihilation operator of an electron at site $n$
with spin $\alpha$.  The Hamiltonian (\ref{spinchain}) is tranformed 
to the Lagrangian
\beeq
L^{(1)}= \sum \Big\{ ~ i c_n^\dagger \dot c_n^{} 
   + \phi_n (c_n^\dagger c_n^{} - 1)    
-{J\over 2} (\chi_n^* \chi_n - \chi_n^{} c_n^\dagger c_{n+1}^{} - 
\chi_n^* c_{n+1}^\dagger c_n^{} ) 
~ \Big\} ~.
\label{spinLagrangian1}
\eneq
$\chi_n$ is an link variable, defined on the link connecting
sites $n$ and $n+1$.  The Lagrangian $L^{(1)}$ has local $U(1)$ gauge
invariance as well.

In a spin chain,
the magnitude of $\chi_n$ is almost frozen, ie.\
  the effective potential for $|\chi_n|=\chi$ has a sharp
minimum at $\chi = 1/\pi$,  the curvature
there being proportional to the lattice volume.  To a very good
approximation, one can write
$\chi_n = (i/\pi) \, e^{ia_0A_n}$ where $a_0$ is the lattice spacing.
$\phi_n$ and $A_n$ are the time and space components of a $U(1)$ gauge
field $A_\mu(x)$.

In an anti-ferromagnetic spin
chain two adjacent sites form one block.  Each block contains four electron
states. The coupling of the gauge field $A_\mu$ in $L^{(1)}$ is spin-blind.
Therefore,  an electron spin  becomes
flavor in the continuum Dirac field, whereas an even-odd site index
becomes a spin index.

\beqn
&&\hskip .4cm c_{a\alpha} 
\hskip .7cm \Longleftrightarrow \hskip .8cm \psi^{(\alpha)}_a \cr
\noalign{\kern 6pt}
\alpha :&& \hskip .3cm \hbox{spin} \hskip 2.1 cm \hbox{flavor} \cr
a :&& \hskip .0cm \hbox{even-odd} \hskip 1.8cm \hbox{spin} 
\eeqn

In the continuum limit $L^{(1)}$ becomes
\beeq
\L^{(2)} = - {1\over 4e^2} \, F_{\mu\nu}^2 
 + \sum_\alpha \psibaralpha i \gamma^\mu (\dd_\mu - iA_\mu)
\psi^{(\alpha)} + {1\over a_0} \, A_0 ~~.
\label{spinLagrangian2}
\eneq
We have added the Maxwell term with the understanding that the limit
$e^2 \go \infty$ is taken at the end.  Light velocity in $\L^{(2)}$
is given by $c = a_0 J/\pi$.

A $s=\onehalf$ spin chain is equivalent to 2-flavor massless QED$_2$ in
the strong coupling limit in a uniform background charge.  After
bosonization 2-flavor QED$_2$ contains two bosons, one with (mass)$^2=
2e^2/\pi$ and the other with a vanishing mass.  In the $e^2\go\infty$ limit
the former decouples.  The latter, the massless boson, is the gapless mode
known in the Bethe ansatz solution. It controles the long-range behavior of
various correlation functions.

\section*{Acknowledgment}
This work was supported in part by  the U.S.\ Department of Energy
under contract DE-AC02-83ER-40105.

\end{document}